\documentclass[a4paper]{article}

\usepackage{INTERSPEECH2022}
\usepackage{enumitem,kantlipsum}
\usepackage[table,xcdraw]{xcolor}
\usepackage{url}
\usepackage{multirow}
\usepackage{enumitem}
\usepackage{xspace}
\newcommand{\MADA}{\texttt{MADA}\xspace}
\newcommand{\twostage}{\texttt{TSP}\xspace}
\newcommand{\adimaclass}{\texttt{AC}\xspace}
\newcommand{\stackclass}{\texttt{SC}\xspace}
\newcommand{\adima}{\texttt{ADIMA}\xspace}

\title{Multilingual and Multimodal Abuse Detection}
\name{Rini Sharon, Heet Shah, Debdoot Mukherjee, Vikram Gupta}
\address{ShareChat, India}
\email{\{rinisharon, heetshah, debdoot, vikramgupta\}@sharechat.co}

\begin{document}

\maketitle
\begin{abstract}
The presence of abusive content on social media platforms is undesirable as it severely impedes healthy and safe social media interactions. While automatic abuse detection has been widely explored in textual domain, audio abuse detection still remains unexplored. In this paper, we attempt abuse detection in conversational audio from a multimodal perspective in a multilingual social media setting. Our key hypothesis is that along with the modelling of audio,
incorporating discriminative information from other modalities can be highly beneficial for this task. Our proposed method, \MADA, explicitly focuses on two modalities other than the audio itself, namely,  the underlying emotions expressed in the abusive audio and the semantic information encapsulated in the corresponding textual form. Observations prove that \MADA demonstrates gains over audio-only approaches on the \adima dataset. We test the proposed approach on 10 different languages and observe consistent gains in the range 0.6\%-5.2\% by leveraging multiple modalities. We also perform extensive ablation experiments for studying the contributions of every modality and observe the best results while leveraging all the modalities together. Additionally, we perform experiments to empirically confirm that there is a strong correlation between underlying emotions and abusive behaviour.\footnote{We will share the code publicly for reproducing our results.}



\end{abstract}

\section{Introduction}
Timely detection and prevention of abusive behaviour on social media is an important problem as multiple users face harassment in the form of targeted personal/communal attacks, which leads to an unpleasant user experience and can cause lasting psychological effects \cite{7275970, o2018social}. With an exponential increase in the use of social media by people from different regional and cultural backgrounds, it is of paramount importance to develop accurate content moderation models for recognizing and filtering offensive interactions in multilingual social media environment. 

Abuse detection in visual~\cite{gao-etal-2020-offensive, alcantara-etal-2020-offensive} and textual~\cite{davidson2017automated, safaya2020kuisail, nobata2016abusive, park2017one} domains has been well explored. However, abuse detection in spoken audio is relatively under explored. In this work, we attempt to identify the presence of abusive words in spoken utterances in a multilingual setting. A two-stage process of transcribing the spoken audio into text using automatic speech recognition (ASR) systems followed by using natural language processing based abuse detection methods designed for text is a plausible approach. While the transcribed text captures the semantic information, it is not able to represent the audio cues like pitch, volume, tone, emotions and so on, which can often play an important role in abuse detection as humans are generally angry, agitated or loud while displaying abusive behaviour \cite{rana2022emotion,plaza2021multi}. 


Considering these issues, we propose a multimodal approach which leverages three important facets of information present in the utterances: a) content present in the audio b) emotions expressed in them and c) textual semantics of the waveform, for robust detection of abuse in spoken audio. We use modality specific models for extracting the features for each modality followed by multimodal fusion and observe substantial improvements by using these modalities in an end-to-end setup for multiple languages. We also observe that only modelling the emotion embeddings showcases competitive performance. These results indicate a strong correlation between the display of abusive behaviour and emotions expressed in the corresponding audio utterances. To the best of our knowledge, our work is the first one to explore the combination of audio, emotions and transcribed text for abuse detection in conversational speech. 
Our major contributions can be summarized as:
\begin{itemize}
    \item We explore a multimodal approach for abuse detection in multilingual spoken text.
    \item We show that explicitly integrating emotions and semantic information present in the transcribed text results in promising state-of-the-art performance. 
    \item We also demonstrate that there exists a strong correlation between emotions and abusive behaviour.
    \item We perform extensive experiments to highlight the contribution of each modality and compare them with a two-stage process (transcription followed by text classification) for abuse detection. 
\end{itemize}


\section{Related Work}

Significant efforts in the domain of abuse detection have been devoted to abusive text \cite{davidson2017automated, safaya2020kuisail}, image \cite{gangwar2017pornography}, video \cite{gao-etal-2020-offensive, alcantara-etal-2020-offensive} and comments \cite{nobata2016abusive, park2017one}. Existing approaches that are employed to perform abuse detection vary from using lexical features together with meta data \cite{koufakou2020hurtbert}, to machine/deep learning models, and their derivatives \cite{kaur2021abusive, jhaveri2022toxicity}. Furthermore, various natural language processing(NLP) approaches involving transfer learning, distillation \cite{wei2021offensive}, and keyword-based protocols have also gained popularity. While audio-based abuse detection is challenging and fairly uncommon, there have been previous attempts to perform isolated limited-vocabulary swear word detection \cite{endah2017automation}, read-out abuse classification \cite{rahut2020bengali} and small-vocabulary keyword spotting \cite{ablaza2014small}. However, these approaches do not capture the overall context as they do not deal with conversational real-world examples and require set vocabulary or templates of abusive words. Moreover, abusive words are ever-evolving and are usually not spoken completely and clearly, making the understanding of the overall context very important for this task.

\begin{figure*}[t]
  \centering
  \includegraphics[width=\linewidth,trim={0 0 0 1cm},clip]{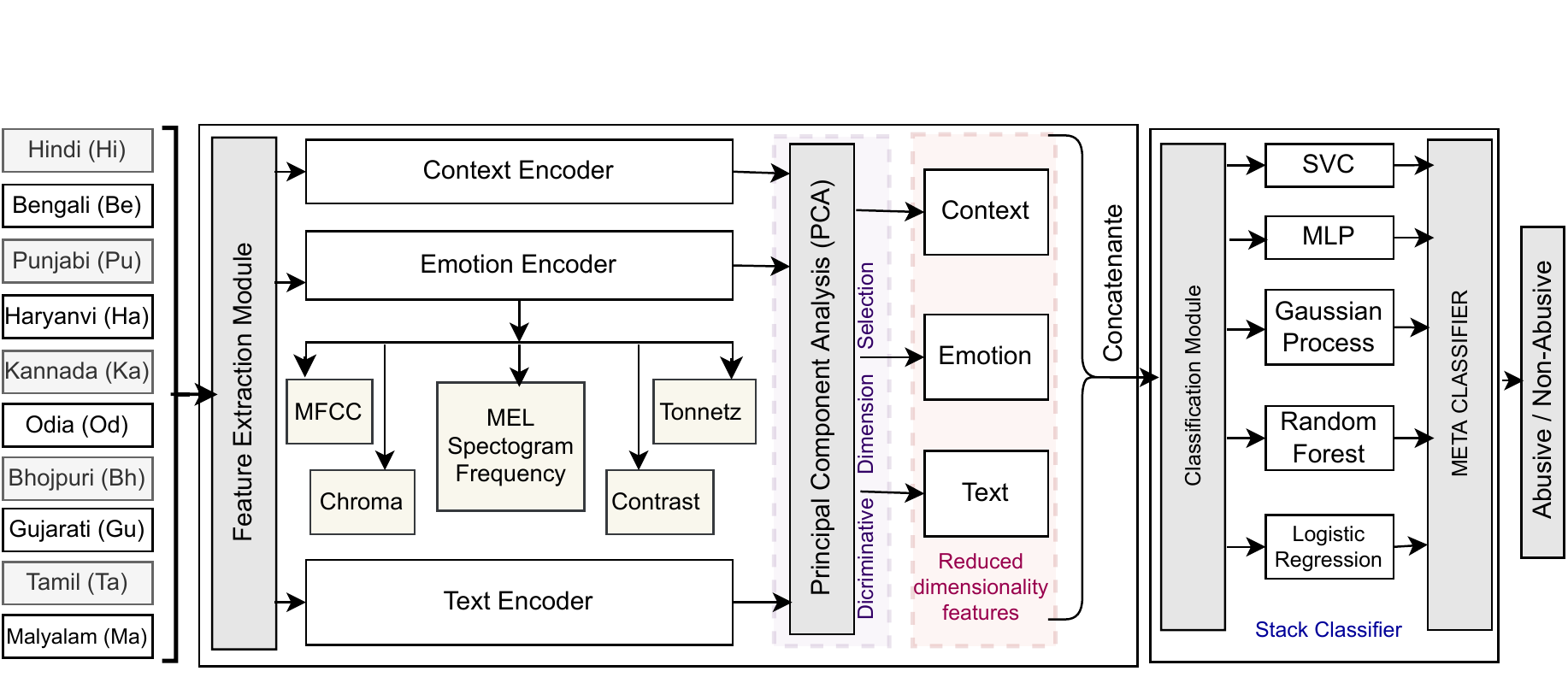}
  \vspace{-5mm}
  \caption{Overall architecture of \MADA}
  \vspace{-3mm}
  \label{fig:effr}
\end{figure*}

Incorporating emotion attributes (which are strongly linked to offensive behaviour in real-life) boosts the performance of abuse detection as opposed to using only the video or text based features as shown in~\cite{rana2022emotion,plaza2021multi, rajamanickam-etal-2020-joint}. Furthermore, research suggests that emotion can be detected with greater accuracy in speech than from its corresponding text \cite{atmaja2019speech, sailunaz2018emotion}, making emotion one of our main choices among modalities. 

Majority of the popular abuse detection approaches in NLP deal with offensive content published via comments in online platforms \cite{jhaveri2022toxicity,nobata2016abusive, safaya2020kuisail}. We therefore leverage the advancements in NLP by transcribing our audio utterances into text. An added advantage while dealing with transcribed data is that intentional textual obfuscations which are challenging to detect \cite{ vidgen2019challenges} such as the use of symbols ``*/\$/." and spelling variations, does not occur. This motivates us to study the text modality as well. To the best of our knowledge, interplay of emotion and textual transcriptions for hate-speech detection has not been explored till date. A closely related work on audio abuse detection is studied in \cite{ghosh2021speech} using annotated English speech from various open-sourced databases. However,~\cite{ghosh2021speech} do not explore the problem from a multimodal and multilingual perspective. 



To fill this gap, in this paper, we explore abuse detection in audio by leveraging audio features alongside modelling emotions and semantics encapsulated in the transcribed text in a multimodal and multilingual setup. We also perform experiments to evaluate the effectiveness of emotions for this task and observe competitive performance by using the emotion representations alone as inputs. 


\section{Task}
\label{sec:task}
We consider the task of classifying audio utterances into abusive and non-abusive categories using different modalities individually and in combination. In this work, audio, emotion and textual features are extracted using modality-specific encoders and fused by concatenation before passing them through the classifier. Formally, let X = $(x_1,...,x_n)$ be the input features fused from multiple modalities of $n$ data instances. Then our output $Y = (y_1,...,y_n)$s are corresponding target labels where $y_i \in \{0, 1\}$, with $0$ indicating non-abusive utterance and $1$ indicating the presence of abuse.


\section{Multimodal Abuse Detection in Audio: \MADA}
\label{sec:multimodal_abuse}
 
In this section, we discuss the overall architecture and core components of \MADA in detail. The following subsections explain in detail the three encoders proposed in our approach for multimodal feature extraction. The features extracted from the encoders are passed through a dimensionality reduction step before performing early fusion across modalities. The fused features are then passed through a binary classifier and trained using binary cross entropy loss. Figure~\ref{fig:effr} shows the overall architectural flow of our approach.

\subsection{Audio Encoder}
\label{sec:texttrans}

For extracting features from the audio, we use wav2vec2 \cite{baevski2020wav2vec, baevski2021unsupervised}, which is a transformer-based model trained in a self-supervised way in a multilingual framework. We consider two pretrained wav2vec2 models, namely, XLSR-53 model (trained using 53 languages which include a few Indic languages) \cite{conneau2020unsupervised} and  CLSRIL-23 (trained using Indic languages) \cite{gupta2021clsril}. We first pass the raw audio files to extract the context embeddings from the fully connected layer and then mean-pool across the temporal dimension to generate the final embeddings.

\subsection{Emotion Encoder}
Similar to previous work \cite{arya2021speech, speechemo}, we extract five features from the audio utterances using the librosa library \cite{mcfee2015librosa}.
\begin{enumerate}[nolistsep]
    \item MFCC: Mel-frequency cepstral coefficients model human perception of frequency and loudness. We extract 40 coefficients here. 
    \item Chroma: Chromagram from a waveform is the representation of the spectral energy with pitch profiling. 
    \item MEL: Mel-scaled power spectrogram is the audio spectrogram with frequencies converted to mel scale. We set the number of mel bands to be generated as 128.
    \item Spectral Contrast: Spectral contrast is the decibel difference between peaks and valleys in the spectrum.
    \item Tonnetz: Tonal centroid features
    arranges sounds according to pitch relationships into interdependent spatial and temporal structures.
\end{enumerate}
These five features are then concatenated together and used as the emotion representation for the respective utterance.

\subsection{Text Encoder}

Firstly, we transcribe the spoken audio to text using pretrained wav2vec models which are finetuned to perform ASR on specific languages of interest (sourced from huggingface \cite{hchadda_huggingface}). We then obtain text embeddings for the transcribed sentence using Sentence-BERT~\cite{reimers-2019-sentence-bert}. We apply the temporal mean-pooling operation on top of the generated word embeddings thereby mapping each input to a 768 dimensional vector space.


\subsection{Feature Fusion}
The feature representations obtained in the previous step are concatenated together and used as the representation of the audio utterance. However, this results in high dimensional feature representations that impede the training process and make it harder to find a good solution space.  To address this, we use Principal Component Analysis (PCA) \cite{jolliffe2016principal} to reduce the dimensions of the feature representation by removing redundant or non-discriminative dimensions. Since PCA is very sensitive to the ranges of the feature values, we apply z-score normalization prior to dimensionality reduction and select the number of PCA components such that the amount of variance to be explained is 95\%. Once these projected features are obtained, they are concatenated to form the final feature vector.


\subsection{Classification Module}

After extracting the features from audio utterance, we train a binary classifier which uses these representations as input and learns to classify them into \textit{abusive} and \textit{non-abusive} categories. We experiment with the classifier architecture used by~\cite{adima} and also propose a stack classifier(SC) architecture\cite{raschka2016mlxtend}. 

\noindent\textbf{ADIMA classifier (\adimaclass):} \adimaclass consists of two fully connected layers($512 \rightarrow256\rightarrow128\rightarrow2$) with ReLU activation. We train using Adam optimizer with learning rate of 0.001, a dropout of 0.1 and a batch size of 16 for 50 epochs. The output is mapped to the probability of the two output classes (\textit{abusive} and \textit{non-abusive}) using softmax layer. 

\noindent\textbf{Stack classifier (\stackclass):} \stackclass is an ensemble learning method and achieves higher predictive performance by combining multiple classification models through a meta-classifier. We use five base classifiers, namely, Gaussian process classifier (kernel “1.0 * RBF(1.0)”), multi-layer perceptron classifier (1 layer, 100 neurons), support vector classifier (linear kernel, regularization parameter C=0.025), random forest classifier (10 trees with max-depth 5) and logistic regression classifier. We train a meta-classifier(logistic regression) over these classifiers for making the final predictions using the stacked predictions of base classifiers as inputs.


\subsection{Two-Stage Process (\twostage)} 


Besides reporting abuse-detection performances of \adima languages using our proposed method, we also compare their performances with the 2-step approach employed in \cite{ghosh2021speech}. In the first step, we use a fine-tuned ASR model for transcribing the audio utterances into text. For each language, we select the ASR model which has been trained for that particular language as quality of the transcription is important for the downstream classification task. In the second step, we use a text-based sequence classifier (Indic-bert) \cite{kakwani2020indicnlpsuite} for classifying the utterance transcriptions as abusive or not. \twostage ($768 \rightarrow 2$) is trained with similar hyper-parameters as the \adima classifier (\adimaclass).


\section{Abusive Behaviour and Emotions}

\begin{figure}[t]
  \centering
  \includegraphics[width=0.75\linewidth]{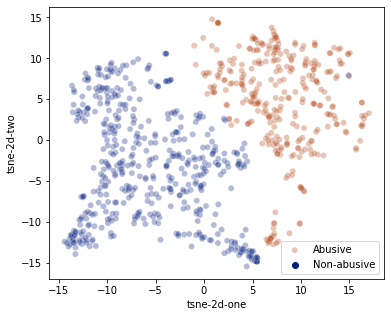}
  \vspace{-3mm}
  \caption{t-SNE visualization of emotion embeddings}
  \vspace{-3mm}
  \label{fig:tsne}
\end{figure}

Abusive behaviour in natural conversations is often accompanied with elevated emotions as humans are generally angry and agitated while displaying abusive behaviour. To validate this hypothesis, we train our models using only emotions as the input feature and compare them with the audio embeddings extracted using the audio encoder. From Table~\ref{tab:example}, we observe that both the features perform competitively for this task which shows that there is correlation between emotions and abusive behavior. In Figure~\ref{fig:tsne}, we show the t-SNE plot of the emotion embeddings for the utterances and observe that the emotion embeddings of abusive and non-abusive words are fairly separable. This also explains the strong performance reported by \adima ~\cite{adima} while using features extracted using a CNN based model trained to perform sound classification instead of human speech understanding. \adima dataset contains audios collected from over 6k users across 10 languages which shows that this correlation holds for multiple users and languages. However, we acknowledge that \adima dataset might have a sampling bias and thus concretely making this claim requires a comprehensive study. 

\begin{table}[t]
\caption{Comparison of abuse classification results between the \adima baselines\cite{adima}, the proposed \MADA and \twostage. For \adima and \MADA, we report the best accuracies. (*) involve XLSR feats and the others CLSRIL feats}
  \label{tab:alllang}
\centering
\begin{tabular}{
>{\columncolor[HTML]{EFEFEF}}c c
>{\columncolor[HTML]{EFEFEF}}c c
>{\columncolor[HTML]{EFEFEF}}c c
>{\columncolor[HTML]{EFEFEF}}c }
\cellcolor[HTML]{EFEFEF} &
  \multicolumn{2}{c}{\cellcolor[HTML]{EFEFEF}\adima} &
  \multicolumn{2}{c}{\cellcolor[HTML]{EFEFEF}\twostage} &
  \multicolumn{2}{c}{\cellcolor[HTML]{EFEFEF}\MADA(Ours)} \\
\multirow{-2}{*}{\cellcolor[HTML]{EFEFEF}\textbf{\begin{tabular}[c]{@{}c@{}}Lang\end{tabular}}} &
  \cellcolor[HTML]{EFEFEF}\textbf{Acc} &
  \textbf{F1} &
  \cellcolor[HTML]{EFEFEF}\textbf{Acc} &
  \textbf{F1} &
  \cellcolor[HTML]{EFEFEF}\textbf{Acc} &
  \textbf{F1} \\
\textbf{Hi} & 79.67          & 79.48                   & 79.0 & 78.49                   & \textbf{84.82}* & \textit{\textbf{83.82}} \\
\textbf{Be} & 81.08*          & 79.46                   & 76.0 & 75.06                   & \textbf{82.43}* & \textit{\textbf{85.58}} \\
\textbf{Pu} & 82.01          & 82.01                   & 78.0 & 77.87                   & \textbf{85.01} & \textit{\textbf{85.09}} \\
\textbf{Ha} & 81.15          & 81.12                   & 68.0 & 68.26                   & \textbf{83.06} & \textit{\textbf{81.87}} \\
\textbf{Ka} & 82.92*          & \textit{\textbf{80.15}} & 58.0 & 56.45                   & \textbf{83.47}* & 79.94                   \\
\textbf{Od} & \textbf{83.29}* & 82.21                   & 82.0 & \textit{\textbf{82.33}} & 82.46          & 77.78                   \\
\textbf{Bh} & 76.48*          & 72.30                   & 73.0 & 73.03                   & \textbf{78.27} & \textit{\textbf{84.24}} \\
\textbf{Gu} & 80.94          & 76.38                   & 69.0 & 68.08                   & \textbf{82.04} & \textit{\textbf{77.62}} \\
\textbf{Ta} & 80.59*          & 75.04                   & 82.0 & \textit{\textbf{82.94}} & \textbf{82.21} & 81.13                   \\
\textbf{Ma} & \textbf{86.29} & \textit{\textbf{83.41}} & 77.0 & 76.77                   & 85.48          & 80.27                  
\end{tabular}
\vspace{-2mm}
\end{table}

\begin{table}[t]
\caption{Ablation study on effect of different modalities (audio, emotion and text) and their combination (all) using concatenation on abuse detection performance (Accuracy (\%))}
\vspace{-1mm}
\label{tab:example}
\centering
\addtolength{\tabcolsep}{0pt}
\begin{tabular}{c c c c c c }
\toprule
{\cellcolor[HTML]{FFFFFF}\textit{\textbf{}}} & {\cellcolor[HTML]{FFFFFF}\textit{\textbf{audio}}} & {\cellcolor[HTML]{FFFFFF}\textit{\textbf{emo}}} & {\cellcolor[HTML]{FFFFFF}\textit{\textbf{text}}} & 
\textit{\textbf{all-AC}} & \textit{\textbf{all-SC}} \\
\midrule
{\color[HTML]{000000} \textit{\textbf{Hi}}} & 79.13 & 78.86 & 78.86 & 83.46 & \textbf{84.82} \\
\textit{\textbf{Ha}} & 79.51 & 78.96 & 69.39 &  82.78 & \textbf{83.06} \\
\textit{\textbf{Ta}} & 80.59 & 81.40 & 77.35 & \textbf{82.21} & \textbf{82.21} \\
{\color[HTML]{000000} \textit{\textbf{Be}}} & 80.81 & 80.54 & 79.19 & 78.91 & \textbf{82.43} \\
{\color[HTML]{000000} \textit{\textbf{Pu}}} & 81.74 & 80.38 & 77.80 & 83.65 & \textbf{85.01}\\
\bottomrule
\end{tabular}
\vspace{-4mm}
\end{table}


\begin{figure}[t]
  \centering
  \vspace{-3mm}
  \includegraphics[width=\linewidth,trim={0 1cm 4.5cm 1.5cm},clip]{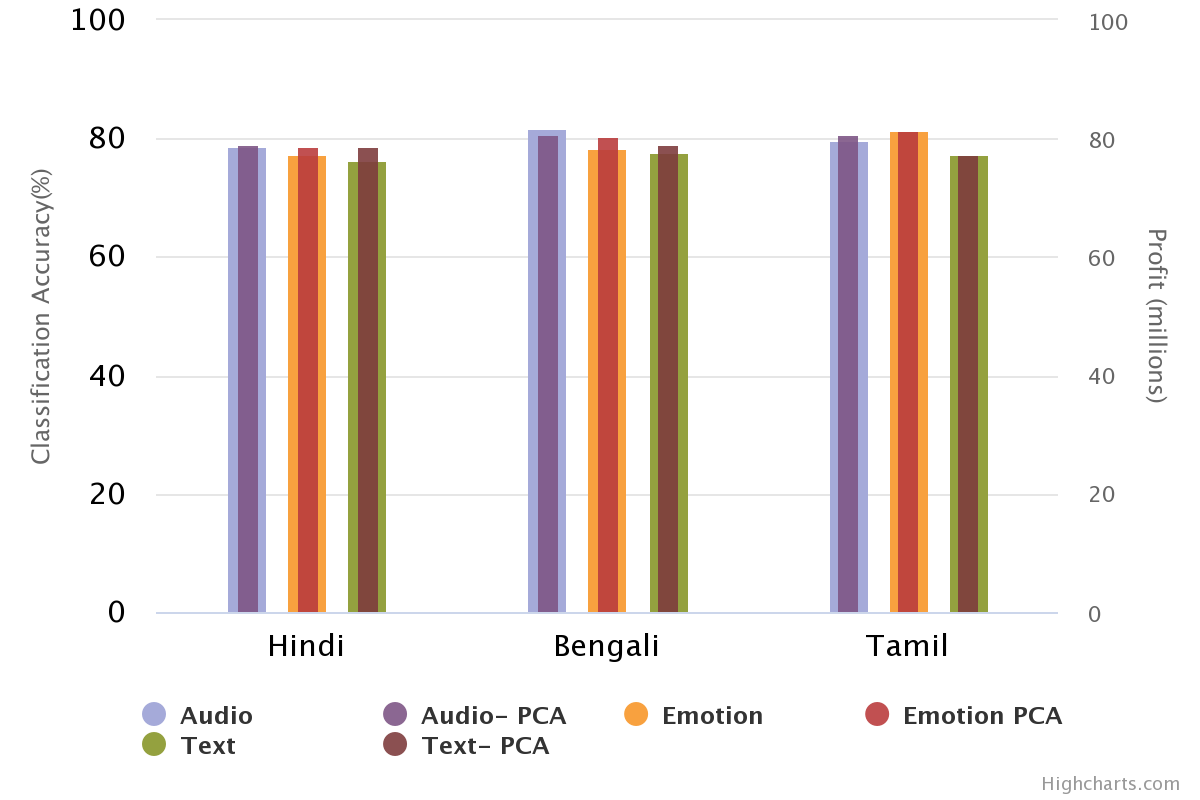}
  \vspace{-4mm}
  \caption{Classification accuracy of \MADA and all modalities post dimensionality reduction. We note that for every modality, dimensionality reduction shows performance gains.}
  \vspace{-5mm}
  \label{fig:pca_fig}
\end{figure}

\section{Dataset and Evaluation}
\label{sec:dataset}

We experiment with Abuse Detection in Multilingual Audio (\adima)~\cite{adima} dataset. \adima is a well balanced multilingual dataset comprising of 10 Indic languages, Hindi (Hi), Bengali (Be), Punjabi (Pu), Haryanvi (Ha) , Kannada (Ka), Odia (Od), Bhojpuri (Bh), Gujarati (Gu), Tamil (Ta) and Malyalam (Ma), with 1200 utterances for each language. Each utterance has been annotated by human annotators as \textit{abusive} or \textit{non-abusive} and is 5-60 seconds long in duration. The utterances containing cuss-words, swear-words, and other forms of explicit abuse are labelled as \textit{abusive}. \adima dataset represents real-life scenarios because the utterances have been sourced from real-life conversations and often contain \textit{environmental noise}. Moreover, poor recording devices or improper placement of the recording device at the user's end further contribute to the loss of clarity. Additionally, the multilingual property of the dataset allows for exploration across multiple \textit{languages, accents and dialects}, specifically for resource-impoverished languages. 

 

\section{Results and Discussion}
\label{sec:resdis}
We train our unimodal and multimodal networks using binary cross entropy loss. We use the train/test splits (70:30 for each language) provided by~\cite{adima} across all the experiments for fair comparison and report accuracy and F1 score on the test set.

\subsection{Classification Results}
 
\label{sssec:subsubhead}
We compare the performance of our proposed method (\MADA) with the baselines reported in ~\cite{adima} and the binary classification results obtained using the two stage process (\twostage) on the \adima dataset. \adima experimented with different backbones and feature processing techniques. In this study, we compare our method with the best results reported by \adima for each language irrespective of the set-up they used. From Table~\ref{tab:alllang}, we observe that our method outperforms the best of \adima results for a majority of the languages (except Odiya and Malyalam) with substantial margin demonstrating the importance of leveraging all the modalities. 

Similarly, we also note that our proposed method improves upon the \twostage method significantly for a majority of the languages. Moreover, the performance of \twostage degrades substantially for some of the languages which could be a result of the inferior quality of ASR models and their transcriptions thereby. Since the ASR models used in \twostage are pretrained models borrowed from different domains, they may not contain instances of abuse words in their vocabulary (out of vocabulary). This therefore may result in substitutions or deletions justifying the dip in evaluation accuracy. Furthermore, day-to-day utterances from Indian languages often contain code-switched temporal portions which make recognition challenging \cite{kapoor2019mind}. Usage of language specific models result in these words not being correctly identified. However, a more detailed study is warranted to establish these justifications conclusively. 

Overall, our proposed method (\MADA) performs the best for majority of the languages with an absolute accuracy gain ranging from \textbf{0.6-5.2\%} (over \adima baselines) and \textbf{0.2-25.5\%} (over \twostage). These results highlight the significance of discriminative contribution by different modalities for this task.

\subsection{Ablation Studies}
\noindent\textbf{Modality} In this section, we study the contributions of different modalities and hyper-parameters for our task using different languages (Hi, Ha, Ta, Be and Pu). For analyzing the significance of modalities, we perform our experiments individually with each modality (audio, emotion and text) and by adding all the different modalities together. From Table~\ref{tab:example}, we can observe that adding the modalities is beneficial and improves the performance significantly for all the languages. For example, the accuracy of Hindi (Hi) improves from \textbf{79.13\%} to \textbf{84.42\%}. Similar trends are also observed for all the other languages.

\noindent\textbf{Correlation between Emotions and Abuse} As discussed in previous section, it is interesting to note that \textit{audio} and \textit{emotion} modality have comparable abuse detection performance as shown in Table~\ref{tab:example}. This observation alludes towards strong correlation between emotions expressed in the audio and abusive behaviour displayed by the speakers. While dealing with emotion based abuse detection, the few false positives that occur may result from cases where the speaker expresses frustration or uses profanities for emphasis or humor, while the false negatives may be caused by missing out on subtler forms of abuse.

\noindent\textbf{Classifier Architecture} We also compare the performance of \adima classifier (\adimaclass) and Stacked classifier (\stackclass) in Table~\ref{tab:example} to isolate the contribution of multimodal fusion and classifier architecture. We note that both classifiers achieve competitive results showcasing that the gains can be attributed to the fusion of the modalities and are classifier-agnostic. 

\noindent\textbf{Dimensionality Reduction}
In Figure~\ref{fig:pca_fig}, we analyze the performance of \MADA on performing dimensionality reduction before training the classifiers for three languages. We note that for all the modalities, dimensionality reduction helps in improving the performance of \MADA. For Hindi, the performance improves by an absolute percentage of \textbf{1.14}, \textbf{1.35} and \textbf{2.44} for audio, emotion and text respectively. Similar patterns are also observed for the other languages as can be seen in Figure~\ref{fig:pca_fig}.

\section{Conclusion}
\label{sec:illust}
Prevention of abusive content is crucial for facilitating safe and healthy interactions. In this work, we explore audio abuse detection from a multimodal perspective in a multilingual setting over 10 Indic languages.  We investigate the significance of modelling the audio, its underlying emotions and the semantic information present in its transcribed text to show gains over approaches which only work with the audio modality. Our detailed ablation experiments evaluate the contribution of every modality and show that we achieve substantial success in classifying abusive occurrences when we leverage information from all the modalities together. This therefore highlights the presence of complementary information in the different modalities. We also compare our methods with two-stage process where the first stage involves transcribing the audio into text and second stage comprises of training a text based classifier over the transcriptions. Our unimodal experiments show that using only emotion features is also competitive for this task which highlights that there is correlation between abusive behaviour and emotions.





\def\UrlBreaks{\do\/\do-}

\bibliographystyle{IEEEtran}

\bibliography{template}


\end{document}